\documentclass[12pt]{iopart}
\usepackage{graphicx}
\usepackage{xcolor}
\usepackage{iopams}
\usepackage{bm}

\begin{document}

\title[AMR and memory effect in bulk systems with extended defects]{Anisotropic magnetoresistance and memory effect \\ in bulk systems with extended defects}

\author{K.\,S.\,Denisov, K.\,A.\,Baryshnikov, P.\,S.\,Alekseev, N.\,S.\,Averkiev}

\address{Ioffe Institute,  Politekhnicheskaya 26, St. Petersburg 194021, Russia}
\ead{barysh.1989@gmail.com}

\begin{abstract}
Memory effects can have a profound impact on the resistivity of semiconductor systems, resulting in giant negative magnetoresistance and MIRO phenomena.
This work opens the discussion of the memory effects in 3D conducting systems featured by the presence of
the extended one-dimensional defects, such as screw dislocations or static charge stripes.
We demonstrate that accounting for the memory effect, that is the capture of electrons on collisionless spiral trajectories winding around extended defects, leads to the strong negative magnetoresistance in case when the external magnetic field direction becomes parallel to the defects axis.
This effect gives rise to a significant magnetoresistance anisotropy already for an isotropic Fermi surface and no spin-orbit effects.
The proposed resistivity feature can be used to detect one-dimensional scattering defects in these systems.
\end{abstract}


\vspace{2pc}

\submitto{\JPCM}
%
\maketitle
%
%

\section{Introduction}

Among a broad family of
the magnetotransport phenomena, a particularly important one concerns the anisotropic magnetoresistance (AMR),
that is the dependence of the longitudinal resistivity on the direction of the applied magnetic field.
AMR can have different microscopic origins and it commonly allows one to analyze rather subtle features of a conductive system.
For instance, the AMR of ferromagnets is generally induced by spin-orbit effects~[1-4],
and it can serve to determine the magnetization easy axis and the exchange anisotropy
of magnetic films~[5-7].  
Alternatively, the AMR
in metallic systems points out at the opened character of the underlying Fermi surface~[8-11], or
even at the topological structure of band states, as in the case of the chiral anomaly in topological semimetals~[12-14].


In this paper, we theoretically predict strong
anisotropic magnetoresistance of a three-dimensional nonmagnetic metal with an isotropic Fermi surface
in case when the electron transport is influenced by the scattering on extended (one-dimensional) defects.
The role of this type of defects can be played by screw dislocations~[15], charged stripes in some dodecaboride crystals~[10] or even artificially made tunnels~[16].
We demonstrate that accounting for the Baskin-Magarill-Entin (BME) mechanism of the memory effect~[17]
leads to the significant enhancement of the AMR in the classically strong range of magnetic fields
even despite that the underlying electron spectrum had isotropic character.

Essentially, the BME-mechanism-based memory effect 
takes into account the appearance of collisionless electron cyclotron trajectories in a random static pattern of short-range impurities.
In particular, this phenomenon is responsible for the strong negative magnetoresistance in 2D conducting films~[18]
as well as for the renormalization of the electron scattering time at ultra-high magnetic fields~[19].
Starting from the pioneering work~[17]
and continuing through Bobylev with coauthors and other works [18,20,21] (see full review of the works in the field in [19]) the mechanisms of reduction of static electrical resistance in strong magnetic fields in 2D materials was theoretically discussed.
Moreover, many other mechanisms for the memory effects have been recently proposed to clarify the understanding of MIRO effects~[22-24].

Importantly, the BME-mechanism-based memory effect is restricted to a two-dimensional electron gas motion, as the cyclotron collisionless electrons
appear only in 2D geometry. The memory effects in 3D samples generally have other microscopic origins,
as firstly considered in~[25,26] they are mostly connected
with multiple returns between scattering events either due to long-ranged and short-ranged scattering potentials.
In this manuscript we extend the topic of BME-based memory effect in bulk metals with residential extended defects (EOD).
The key finding of our work is that
in case when the external magnetic field is directed along EOD axis
the electron motion in the plane perpendicular to EOD
mimics the two-dimensional geometry,
thus allowing the electrons to be captured on the spiral trajectories which winds around EOD without experiencing any collisions.
Similarly to the two-dimensional systems~[18],
this effect leads to the significant negative magnetoresistance at classically strong magnetic fields.
Naturally, the enhanced magnetoresistance exists only in one particular geometry and it remains absent for other directions of magnetic field,
thus giving rise to the strong AMR. Since this effect is the direct consequence of scattering processes on EOD, the AMR constitutes a purely electrical tool to evidence the presence of EOD in a bulk system.

The paper is organized as follows.
In Section~\ref{Sec_Magnetoresistance}
we provide a general discussion of the magnetoresistance anisotropy in EOD-based systems
with isotropic Fermi surface and no spin-orbit effects.
In Section~\ref{Sec_Electron_kinetics}
we describe the electron kinetics 
in materials with EOD-scatterers.
In this section we focus on the Boltzmann equation approximation and neglect any memory effects.
In Section~\ref{Sec_Memory_effect} we develop our original approach to the memory effect description in 3D materials with EODs, which is based on the direct calculation of the electron velocity correlation functions with account for the classical electron trajectories.
In Section~\ref{Sec_Results_and_Discussion} we discuss the obtained results,
and we also study the effect from a deviation of the magnetic field direction from the direction of EODs on the memory effect and discuss other possible memory effects in the considered system.
\section{Magnetoresistance}\label{Sec_Magnetoresistance}
\subsection{General consideration}
We consider a bulk metallic system with isotropic electron Fermi surface.
The system is randomly filled by extended 1D defects directed along $z$ axis, see Fig~\ref{fig:model}.
The presence of co-aligned EOD lowers the system symmetry by selecting $z$ axis and, thus, makes the media effectively anisotropic.
The magnetoresistance,
however, does not appear simply due to the anisotropy of an effective media.
Indeed, the equation of motion for an average drift carrier velocity ${\bf v}$ upon the external electric ${\bf E}$ and magnetic ${\bf B}$ fields
with the anisotropic friction term is given by
\begin{equation}
\frac{d {\bf v}}{dt}=  \frac{e}{m} {\bf E} + \frac{e}{mc} \left[ {\bf v} \times {\bf B} \right] - \hat{\Gamma} \cdot {\bf v},
\end{equation}
here $\hat{\Gamma} = {\rm diag}(\tau_0, \tau_0, \tau_z)$ is the anisotropic tensor of the scattering times, $\tau_z \neq \tau_0$.
The tensor of resistivity is determined by the emergent electric field ${\bf E} = \hat{\rho} \cdot {\bf j}$,
which appears due to the applied electric current density ${\bf j} = n e {\bf v}$. In the steady state
\begin{equation}
\label{eq_NO-M}
    {\bf E} = \frac{m}{n e^2} \hat{\Gamma} \cdot {\bf j} - R_H \left[ {\bf j} \times {\bf B} \right]   \equiv \hat{\rho} \cdot {\bf j},
\end{equation}
where $R_H = (nec)^{-1}$ is the Hall coefficient.
It follows from (\ref{eq_NO-M}) that the ${\bf B}$-dependent term contributes to
the $\alpha$ component of ${\bf E}$ only via the perpendicular components of the electrical current.
Thus,
in such a simplified model
the magnetic field affects only the Hall resistivity and has no effect on the longitudinal elements of $\hat{\rho}$, hence no magnetoresistance.

A general mechanism giving rise to the magnetoresistance
concerns the spreading of the electron drift velocity.
Generally speaking, instead of using the basic relation ${\bf j} = n e {\bf v}$
we should firstly calculate the total electric conductivity by summing over different electrons, 
and only after that should we inverse the conductivity tensor.
For instance, the energy dependence of the electron scattering rates $\Gamma(\varepsilon)$
leads to the non-uniform velocity distribution which directly results in the magnetoresistance~[27].
This mechanisms leads to the positive quadratic magnetoresistance at weak magnetic field and it becomes suppressed in the strong limit.

In our paper we focus on several microscopic mechanisms of the magnetoresistance, which are specifically relevant for the considered setting
and which stem from the spreading of the electron drift velocity due to the electron scattering on EOD.
Firstly, we are going to consider the electron kinetics with account for the angular dependence of
the electron scattering rates $\Gamma({\bf v})$ on EOD,
this mechanism is mostly relevant at weak magnetic field.
Secondly, we investigate the Baskin-Magarill-Entin mechanism of the memory effect due to the formation of
collision-free spiral electron trajectories winding around EOD in case when the magnetic field is directed along their direction.
This mechanism is highly effective at strong magnetic fields and it leads to a significant increase of AMR.

\subsection{Model}

\begin{figure}[t]
\centerline{\includegraphics[width=1\textwidth]{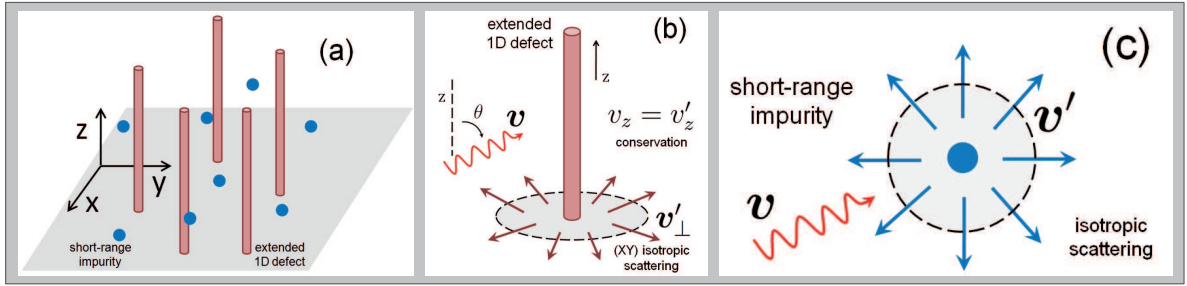}}
\caption{Scattering on short-range impurities and extended 1D defects.}
\label{fig:model}
\end{figure}

Let us describe the considered model in more detail (see Fig.~\ref{fig:model}).
We assume that the bulk system contains two types of elastic scatterers, namely the EOD directed along $z$-axis and additional short-range impurities.
The electron scattering on short-range impurities is isotropic
(the electron velocity directions before and after the scattering event are totally uncorrelated),
it is described by a single parameter, which is the 
scattering time $\tau_i$.
The electron scattering on EOD is more complex, see Fig.~\ref{fig:model}.
In view of the transnational invariance along z axis
the electron momentum 
$p_z = p_F \cos{\theta}$ is conserved upon the scattering on EOD,
here $p_F$ is the Fermi momentum.
Also, the scattering on EOD with respect to
the electron momentum projection ${\bf p}_{\perp} = p_x {\bf e}_x + p_y {\bf e}_y$ onto ($xy$) plane, is assumed to be isotropic.

We focus on 3D degenerate electron gas and present the conductivity tensor in the following way
\begin{equation}
    \hat{\sigma}({\bf B}) = e^2 \nu_F \hat{\mathcal{D}}({\bf B}),
\end{equation}
where $\nu_F$ and $\mathcal{D}_{\alpha \beta}$ are the electron density of states and the diffusion tensor respectively, both are taken at the Fermi energy.
The resistivity tensor is obtained by inverting $\hat{\rho} = \hat{\sigma}^{-1}$
\begin{equation}
\label{eq_res0}
    \hat{\rho}({\bf B}) = \frac{1}{e^2 \nu_F} \hat{\mathcal{D}}^{-1}({\bf B}).
\end{equation}
In what follows
we are going to consider the anisotropy of $\hat{\rho}$ in the static limit ($\omega \to 0$) for two geometries
with respect to the mutual orientation of the applied magnetic field and the EOD axis:
${\bf B} \parallel {\bf e}_z$ and ${\bf B} \parallel {\bf e}_y$.
The anisotropy of magnetoresistance
is described by
a difference of resistivities in these two types of the considered geometries
\begin{equation}
\label{eq_DeltaAM}
    \Delta_{AM}(B) \equiv \frac{\rho_{xx}({\bf B} \parallel {\bf e}_y ) - \rho_{xx}({\bf B} \parallel {\bf e}_z )}{\rho_{xx}(B=0)}.
\end{equation}
The AMR 
within the kinetic equation approximation
due to the $\theta$-dependence of EOD scattering rates
is described in Sec.~\ref{Sec_Electron_kinetics}.
The strong enhancement of $\Delta_{AM}$  due to the memory effect is further considered in Sec.~\ref{Sec_Memory_effect}.

\section{Electron kinetics}
\label{Sec_Electron_kinetics}




Let us consider the electron transport on the basis of the Boltzmann kinetic equation.
The electron distribution function $f_{\bf p} = f_{\bf p}^0 + \delta f_{\bf p}$, which contains the equilibrium $f_{\bf p}^0$ and nonequilibirum $\delta f_{\bf p}$ parts,
satisfies
\begin{equation}
\label{eq-kinetic}
    e ({\bf E} \cdot {\bf v} ) \frac{\partial f_{\bf p}^0}{\partial \varepsilon} + \frac{e}{c} \left[ {\bf v} \times {\bf B} \right] \cdot
    \frac{\partial f_{\bf p}}{\partial {\bf p}} = - \frac{\delta f_{\bf p}}{\tau_i}
    - \frac{\delta f_{\bf p} - \langle \delta f_{\bf p} \rangle_\varphi}{\tau_s},
\end{equation}
where ${\bf E},{\bf B}$ are the external electric and magnetic fields,
${\bf v} = {\bf p}/m$ is the electron velocity,
and we presented the collision integral as two terms.
The first one corresponds to the isotropic electron scattering on short-range impurities described by the 
scattering time $\tau_i$.
The second term describes the electron scattering on EOD with the magnitude of rates controlled by the parameter $\tau_s$,
here $\langle \dots \rangle_\varphi$ stands for the average over the polar angle $\varphi$ in $(xy)$ plane perpendicular to EOD axis.
This form of the collision integral is the direct consequence of the implied
in-plane isotropic model for the electron scattering on EOD
and it does not lead to the relaxation of $\varphi$-independent parts of the distribution function.
In particular, it preserves
$v_z = v_F \cos{\theta}$ component of the electron drift velocity.
We justify this expression and connect the scattering time $\tau_s$ with microscopic parameters of EOD in the \ref{App-CollisionIntegral}.

It is important to note that depending on the microscopic model the time $\tau_s$
can become dependent on the azimuthal angle $\theta$ of the electron velocity. 
For instance, one can estimate the scattering time on extended defects as follows: $\tau_s^{-1} = n_s v_\perp \sigma_s$,
where $n_s, \sigma_s$ are the sheet density and the scattering cross-section of EOD,
and $v_\perp = v_F \sin{\theta}$ is the electron velocity component in $(xy)$ plane.
Assuming the $\theta$-independent value of $\sigma_s$ we get
the $\theta$-dependent scattering time $\tau_s^{-1}(\theta) \propto \sin{\theta}$,
which is due to the dependence of the electron incident flux intensity on its velocity in ($xy$) plane.
In particular, this model takes into account that the electrons moving strictly parallel to EOD and having $v_z = v_F$ are not scattered by the EOD.
Naturally, the microscopic structure of the electron scattering on EOD is important for the appearance of the magnetoresistance.
We also note that an angular dependence of the scattering rates was previously implemented for the description of the paramagnetic gas kinetics~[28].

The general form of $\delta f_{\bf p}$
solving the kinetic equation can be written in the following way
\begin{equation}
    \delta f_{\bf p} =  e \left( - \frac{\partial f_{\bf p}^0}{\partial \varepsilon}\right)
    {\bf E} \cdot \left( \hat{\mathcal{T}}_{{\bf B}} {\bf v} \right),
    \qquad
    \mathcal{D}_{\alpha \beta} = \Bigl\langle
    v_{\alpha} ( \hat{\mathcal{T}}_{{\bf B}} {\bf v})_{\beta}
    \Bigr\rangle,
\end{equation}
where $\mathcal{D}_{\alpha \beta}$ are the corresponding diffusion coefficients, 
the average goes over the Fermi momentum direction ${\bf n}$,
the notation $\hat{\mathcal{T}}_{\bf B}$ stands for the operator acting on the electron velocity ${\bf v}$, it depends both
on the scattering times and on the direction of the applied magnetic field.
The explicit form of $\hat{\mathcal{T}}_{\bf B}$ and the corresponding expressions for the diffusion tensor depend crucially on the considered geometry.

At zero external magnetic field $B=0$
the action of $\hat{\mathcal{T}}_{\bf B}$ operator is reduced to the multiplication of the velocity by the corresponding scattering time
$\hat{\mathcal{T}}_{\bf B} {\bf v} = (\tau_0 v_x, \tau_0 v_y, \tau_i v_z)$,
where we introduced the overall scattering time
\begin{equation}
\label{eq_TAU0}
    \tau_0^{-1} = \tau_i^{-1} + \tau_s^{-1},
\end{equation}
which is responsible for the relaxation of ($xy$) velocity components.
The diffusion coefficients
and the components of the resistivity tensor
at $B=0$ read as 
\begin{eqnarray}
    & \mathcal{D}_{xx}=\mathcal{D}_{yy} = \langle v_{x}^2 \tau_0 \rangle,
    \quad
    \mathcal{D}_{zz} = \langle v_{z}^2 \tau_i \rangle,
    \quad
    \mathcal{D}_{xy}=\mathcal{D}_{xz}=0,
    \\
    & \rho_{xx} = \rho_{yy} = \frac{1}{e^2 \nu_F \mathcal{D}_{xx}} ,
    \quad
    \rho_{zz}  = \frac{1}{e^2 \nu_F \mathcal{D}_{zz}},
    \quad
    \rho_{xy}=\rho_{xz}=0.
\end{eqnarray}
Naturally, the resistivity is anisotropic $\rho_{xx} \neq \rho_{zz}$,
which is due to the different relaxation times $\tau_i \neq \tau_0$.
We note that for the $\theta$-dependent model the time $\tau_0(\theta)$ also depends on the velocity projection onto EOD axis,
so the average over the velocity directions ${\bf n}$ in formulas for $\mathcal{D}_{\alpha \beta}$ is sensitive to its particular form.

The solution of the kinetic equation at finite magnetic field directed parallel to EOD
(${\bf B} \parallel {\bf e}_z$)
can be obtained analytically for an arbitrary $\tau_s(\theta)$ model.
In this case the distribution function mimics the two-dimensional geometry and it can be presented in the following form
$\delta f_{\bf p} = A(\theta) \cos{\varphi} +  C(\theta) \sin{\varphi}$, where $A,C$ are some functions of $\theta$-angle.
The corresponding diffusion coefficients are given by
\begin{equation}
\mathcal{D} \equiv \mathcal{D}_{xx} + i \mathcal{D}_{yx} =
\Bigl\langle
\frac{v_x^2 \tau_0 }{1- i \omega_c \tau_0}
\Bigr\rangle,
\quad
   \mathcal{D}_{zz} = \langle v_{z}^2 \tau_i \rangle.
   \label{eq_D-A-Kin}
\end{equation}
We note that $\mathcal{D}_{zz}$ coincides with its value at zero magnetic field; in this geometry ${\bf B}$ does not affect the electron motion along EOD.
The in-plane coefficients, on the contrary, are modified accounting for the cyclotron motion in $(xy)$ plane.
For the $\theta$-dependent model $\tau_s(\theta)$ it is the account for the angular averaging in $\mathcal{D}$ that leads to the magnetoresistance.

The solution of the kinetic equation for the geometry with ${\bf B} \parallel {\bf e}_y$ is much more complicated and it essentially depends on the particular $\tau_s(\theta)$ function.
In fact, the analytical solution can be obtained only for the simplest $\theta$-independent model. In this case the diffusion coefficients have the following form
\begin{eqnarray}
\mathcal{D}_{zz} =
    \frac{1}{3}
    \frac{v_F^2 \tau_i}{1+ \omega_c^2 \tau_0 \tau_i},
    \quad
\mathcal{D}_{xx} =
 \frac{1}{3}
\frac{v_F^2 \tau_0}{1+ \omega_c^2 \tau_0 \tau_i},
\\
\mathcal{D}_{zx} = - \mathcal{D}_{xz} =
 \frac{1}{3}
\frac{v_F^2 \omega_c \tau_0 \tau_i}{1+ \omega_c^2 \tau_0 \tau_i}.
\end{eqnarray}
In order to account for some $\theta$-dependence of $\tau_s(\theta)$ and to describe the associated magnetoresistance
we made a numerical calculations of the Boltzmann equation; this case is described in details in the \ref{App-NumSol-Boltzmann}.


\begin{figure}[t]
\centerline{\includegraphics[scale=0.8]{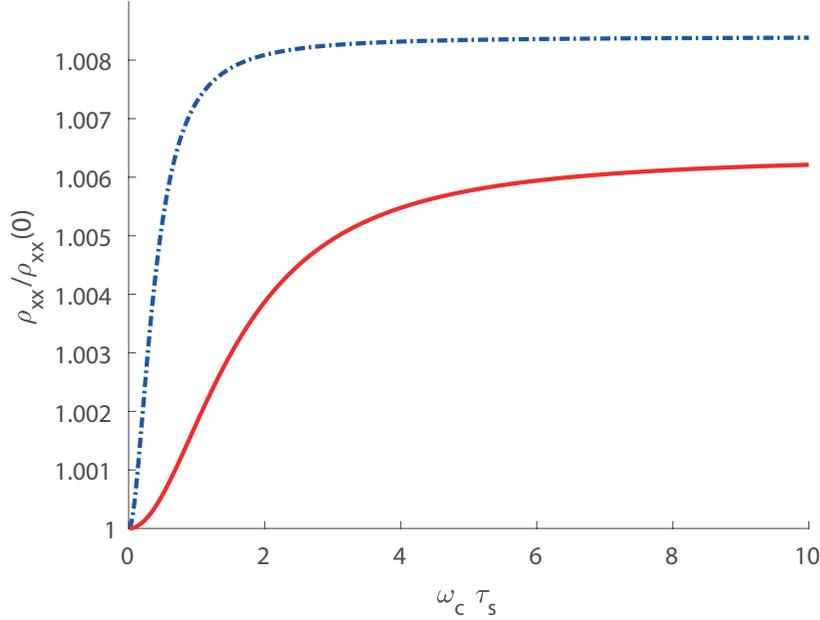}}
\caption{The dependence $\rho_{xx}(B)$
for $B || z$ (red solid line) and $B || y$  \\ (blue dash-dotted line) geometries in $\theta$-dependent model of $\tau_s$,
the \\ ratio $\tau_i/\tau_s^0 =1$.
}
\label{fig:rhoBzBy}
\end{figure}

We proceed with analyzing the anisotropic magnetoresistance due to the $\theta$-dependent electron scattering on the extended defects.
In Fig.~\ref{fig:rhoBzBy} we present the dependence of $\rho_{xx}$ 
on the magnetic field magnitude for two geometries (${\bf B} \parallel  {\bf e}_z$ and ${\bf B} \parallel  {\bf e}_y$); the EOD scattering time was modeled $1/\tau_s = \sin{\theta}/\tau_s^0$ as due to the velocity independent scattering cross-section.
The obtained magnetoresistance is positive for all directions of the magnetic field.
It demonstrates the initial increase for $\omega_c \tau_s \lesssim 2$ with the subsequent saturation in the strong limit $\omega_c \tau_s \gg 1$.
The overall effect remains small, the magnetoresistance in the saturation regime barely approaches $1\%$.
Importantly, we draw the attention to the appearance of the magnetoresistance anisotropy: as follows from Fig.~\ref{fig:rhoBzBy} the resistivity in two geometries indeed differs by several percents.







\section{Memory effect}
\label{Sec_Memory_effect}

\subsection{General consideration}


In this section we investigate  the memory effect based mechanism for the anisotropic magnetoresistance of 3D system with extended defects. 
We demonstrate that 
accounting for the classical collisionless trajectories, which naturally appear when the magnetic field is directed along EOD,
leads to the significant enhancement of AMR at strong magnetic fields.
This effect is closely related to the classical mechanism of the negative magnetoresistance in two-dimensional systems with static disorder due to the Baskin-Magarill-Entin memory effect~[18].
Let us illustrate the latter in more detail.
The close inspection of possible electron 2D trajectories in a given static disorder upon the magnetic field perpendicular to the motion plane
reveals that 
there exist two groups of electrons labeled as the "wandering" and  "circling".
The "wandering" electrons experience occasional scattering and thus behave similarly to the Drude-like electron gas
(at least while $\omega_c$
remains not large enough neither to induce
the localization~[18] the Shubnikov De Haas oscillations),
they contribute both to the longitudinal and transverse electric conductivities.
The "circling" electrons, on the contrary, move along the collision-free circles,
which are inevitably presented in case of static distribution of scalar impurities.
Importantly, the "circling" electrons are deflected by the crossed electric and magnetic fields only in the perpendicular to ${\bf E}$ direction,
thus contributing entirely to the Hall conductivity. 
These two electron groups thus are effectively described by different drift velocities, which
leads to the magnetoresistivity that increases at strong magnetic field ($\omega_c \tau \gg 1$) and has an essentially negative value, hence the negative magnetoresistance.

We draw the attention that
a bulk systems with 3D electrons moving
in a static pattern of the extended defects which are co-aligned with the external magnetic field
naturally imitates the physics of the memory effect in two-dimensions.
Indeed, 
the scattering of electrons by EOD does not lead to the change of $v_z = v_F \cos{\theta}$ velocity component.
Thus the electrons with a fixed projection $v_z(\theta)$ 
can be viewed effectively as a two-dimensional electron gas having the Fermi velocity $v_{\perp} = v_F \sin{\theta}$ and
experiencing the two-dimensional diffusion in ($xy$) plane at $B=0$.
The nonzero magnetic field applied along EOD does not change the magnitude of an effective Fermi velocity $v_{\perp}$, thus
each two-dimensional slice of $\theta$-electrons becomes inevitably separated onto
the "wandering" electrons, which continue to experience the 2D diffusion in ($xy$) plane, and "circling" electrons, which become captured on EOD collisions-free circles. The actual trajectory of these "circling" electrons is, however, the 3D spiral line winding
around EOD along its axis.
We conclude that for each group of electrons with different $v_\perp$
the memory effect due to classical collisionless trajectories gives rise to the negative magnetoresistance.
As a result, the overall resistivity in $(xy)$ plane becomes $B$-dependent and exhibits the strong decrease at large magnetic fields.
Importantly, this effect
can exist only in a particular geometry of ${\bf B} \parallel {\bf e}_z$;
for other orientations of ${\bf B}$ the electron velocity component $v_z$ varies in time, thus suppressing the role of
the static correlations between EOD positions.
Thus the BME-mechanism of the memory effect in bulk systems with static EOD pattern manifests itself as an enhanced anisotropic magnetoresistance.

\subsection{Correlation function technique}

Let us proceed with describing the memory effect in more detail.
The memory effects arise from the electron motion along some classical trajectories whose evolution is fully determined by the initial conditions and, thus,
appears to be free of stochastic changes.
Such a deterministic behavior of an electron motion naturally violates the
treating of the electron kinetics purely in terms of the kinetic equation (\ref{eq-kinetic}).
One common approach
to account for the deterministic trajectories
is to artificially modify the collision integral in the kinetic equation to include possible returns to the same scatterers or non-scattering trajectories in consideration~[20-22].

In this paper we are going to consider this problem based on the correlation function technique. 
According to the fluctuation-dissipation relation, the diffusion coefficients in the static limit ($\omega \to 0$) can be expressed as the time integral
\begin{equation}
\label{eq_D-CFT}
    \mathcal{D}_{\alpha \beta}  = \int\limits_0^{\infty} dt \langle v_{\alpha}(t) v_{\beta}(0) \rangle_{eq}
\end{equation}
over the electron velocity correlation function $ \langle v_{\alpha}(t) v_{\beta}(0) \rangle_{eq}$ in the thermodynamic equilibrium.
In order to prepare the ground for the description of the memory effect, 
we firstly re-consider the computation of the diffusion tensor in the kinetic equation approximation.
For ${\bf B} \parallel {\bf e}_z$ 
the electron trajectory consists of the spiral lines segments
featured by the deterministic cyclotron rotation in ($xy$) plane (and the uniform motion along $z$-axis), which are
interrupted by the occasional scattering.
Importantly,
the kinetic equation approximation signifies that the probability density regulating the location of kinks due to scattering events
on a trajectory is determined by the Poisson stochastic process.
The latter suggests that the probability of an electron to be scattered remains constant in time.
Naturally, assuming the independent Poisson processes both for the scattering on short-range impurities and EOD,
we present the probability density for being scattered for the first time at the interval $(t,t+dt)$ either by short-range impurities $P_i(t)$ or by
EOD $P_s(t)$ in the following way:
\begin{equation}
\label{eq:Poisson}
    P_i(t) = \frac{1}{\tau_i} e^{-t/\tau_0},
    \qquad
    P_s(t) = \frac{1}{\tau_s} e^{-t/\tau_0},
\end{equation}
here $\tau_i,\tau_s$ are the corresponding scattering times,
and $\tau_0^{-1} = \tau_i^{-1} + \tau_s^{-1}$ is an overall scattering time, which accounts for both scattering events.
Naturally, the parameters ($\tau_i,\tau_s,\tau_0$) are essentially the same that we used in the collision integral in (\ref{eq-kinetic}).
Moreover, since the electron $v_z$ component remains constant between collisions,
the formula for $P_s$ can be also applied for the $\theta$-dependent models.
The overall density $P_0 = P_i + P_s$ describes the probability of being scattered for the first time disregard to the particular type of the scattering event.

We calculate the diffusion coefficients by
summing separately the contributions to the time integral (\ref{eq_D-CFT}) from different segments of the electron trajectories.
We also take into account the specifics of our model, which is that
the scattering on short-range impurities is isotropic
and the scattering on EOD is such that it preserves the correlation of only $v_z$ velocity component.
We present here the consideration of the ($xy$) plane diffusion coefficients, the calculation of $\mathcal{D}_{zz}$ is given in the \ref{App-DiffCoef}.
The correlations of ($xy$) velocity components are destroyed
already after the very first collision disregard to the type of scatterer.
Thus the only nonzero contribution to $\mathcal{D}_{xx}, \mathcal{D}_{xy}$ stems from the very first segment of all trajectories that collects the correlations before the first scattering event
\begin{equation}
\label{eq_Dif-XX}
\mathcal{D} \equiv \mathcal{D}_{xx} + i \mathcal{D}_{yx} =
\Bigl\langle
\int\limits_0^{\infty} P_0(\tau_1)  d\tau_1
\int\limits_0^{\tau_1} dt \cdot e^{i \omega_c t} v_x^2(0)
\Bigr\rangle
= \Bigl\langle
\frac{v_x^2 \tau_0 }{1- i \omega_c \tau_0}
\Bigr\rangle,
\end{equation}
where we explicitly took into account the time evolution of an electron velocity due to the cyclotron motion, that means $v_x(t) = v_F \sin{\theta} \cos{(\omega_c t + \varphi_0)}$, $v_y(t) = v_F \sin{\theta} \sin{(\omega_c t + \varphi_0)}$, and the average goes over the initial velocity direction $(\theta,\varphi_0)$; this also implies $\langle v_y(0) v_x(0) \rangle = 0$.
Naturally, we obtained essentially the same expressions for the diffusion coefficient ${\mathcal{D}}$ as we did using the kinetic equation approach, see (\ref{eq_D-A-Kin}).




Let us discuss
the modifications we should make in this calculation in order to take into account the existence of the "circling" electrons.
The integration in (\ref{eq_Dif-XX}) can be presented in the following way
\begin{equation}
    \label{eq_Dif-Mem}
\mathcal{D}
=\Bigl\langle
\int\limits_0^{T_c} P_0(\tau_1)  d\tau_1
\int\limits_0^{\tau_1} dt \cdot e^{i \omega_c t} v_x^2
\Bigr\rangle
+ \Bigl\langle
\int\limits_{T_c}^{\infty} P_0(\tau_1)  d\tau_1
\int\limits_{T_c}^{\tau_1} dt \cdot e^{i \omega_c t} v_x^2
\Bigr\rangle,
\end{equation}
where we divided the integral over the time of the first scattering event $\tau_1$
over two parts, before and after the first cyclotron period $T_c = 2\pi/\omega_c$ is finished.
In fact, one should identify the "circling" trajectory as such,
for which the electron has not yet experienced a scattering after one full cyclotron period $T_c$ is completed.
Indeed, if no short-range impurities were additionally presented,
the fact that an electron have succeeded to accomplish its cyclotron circle without experiencing the collisions with any EOD would suggest
that its cyclic motion is guaranteed to linger forever.
The relative number of so-defined "circling" trajectories $\mathcal{P}_\theta$ for electrons with fixed $\theta$ is thus simply given by
\begin{equation}
    \mathcal{P}_\theta(T_c) = \int\limits_{T_c}^{\infty} P_0(\tau) d\tau = e^{- 2\pi/\omega_c \tau_0}.
\end{equation}
The first term in (\ref{eq_Dif-Mem}) collects the contributions from all electrons
which did experience the collision at $t\le T_c$, those are actually the
"wandering" electrons and we should keep this term as it is.
The second term in (\ref{eq_Dif-Mem}) in its current form describes
the stochastic process of an electron scattering for times greater than $t>T_c$.
Naturally, the inclusion of the memory effect suggests to modify the second term in an appropriate way so that it
excludes the scattering events associated with EOD. 
To do this we note that
for times satisfying $\tau_1 > T_c$
the stochastic probability density  $P_0(\tau_1)$ entering the second integral can be interpreted as
$P_0(\tau_1 > T_c) = \tilde{P}(\tau_1| T_c) \cdot \mathcal{P}_\theta(T_c)$, where
$\tilde{P}(\tau_1| T_c)$ is the conditional probability that an electron will experience some scattering event at time $\tau_1$
provided that it was not scattered for as long as $T_c$ starting from $t=0$; the quantity $\mathcal{P}_{\theta}(T_c)$
here counts the overall probability
that $\theta$-electrons were not scattered by the time $T_c$. 
We further express the diffusion coefficient in the following way
\begin{equation}
    \mathcal{D} =
    \Bigl\langle v_x^2 \tau_0 \cdot
\frac{1 - \mathcal{P}_\theta }{1 - i \omega_c \tau_0}
\Bigr\rangle
+ \Bigl\langle v_x^2 \mathcal{P}_\theta
\int\limits_{T_c}^{\infty} \tilde{P}(\tau_1| T_c)  d\tau_1
\int\limits_{T_c}^{\tau_1} dt \cdot e^{i \omega_c t}
\Bigr\rangle.
\end{equation}
For the kinetic equation approximation the conditional probability entering in the second term
is given by $\tilde{P}(\tau_1| T_c) = \tau_0^{-1} e^{-(\tau_1 - T_c)/\tau_0}$,
thus it takes into account both the short-range impurities and EOD.
In order to exclude the subsequent scattering on EOD for $\tau_1>T_c$,
it is natural to replace the Poisson conditional probability density
by $\tilde{P}'(\tau_1| T_c) = \tau_i^{-1} e^{-(\tau_1 - T_c)/\tau_i}$,
which now takes into account
that at times $\tau_1>T_c$ an electron can be scattered only by short-range impurities, hence the time $\tau_i$ instead of $\tau_0$.
Making this assumption we get the following expression for the diffusion coefficients
\begin{equation}
\label{eq_D-memory}
    \mathcal{D} =
\Bigl\langle v_x^2 \tau_0
\frac{1 - \mathcal{P}_\theta}{1 - i \omega_c \tau_0}
\Bigr\rangle
+ \langle
v_x^2 \mathcal{P}_\theta
\rangle
\frac{\tau_i}{1 - i \omega_c \tau_i}.
\end{equation}
The first and second terms here can be approximately attributed to the "wandering" and "circling" electrons, respectively.
It is now quite clear that these two groups of electrons
cannot be described by the same drift velocity as their contributions to the diffusion coefficients are completely different;
the parameter discriminating between them is $\mathcal{P}_\theta$.
This leads to the magnetoresistance featured by the $B$-dependence similar to the classical negative magnetoresistance of two-dimensional electron gas;
its behavior is considered in detail in the next section.

\section{Results and Discussion}
\label{Sec_Results_and_Discussion}

\subsection{Anisotropic magnetoresistance: calculation and analysis}

We proceed with discussing the anisotropic magnetoresistance emerging due to the memory effect. 
We note that in order to describe $(xy)$ components of the resistivity tensor and, thus, to analyze the emerging magnetoresistance $\rho_{xx}(B)$
for ${\bf B} \parallel {\bf e}_z$ it is sufficient to evaluate a single parameter $\mathcal{D}$. 
Indeed, as soon as $\mathcal{D}_{zx} = \mathcal{D}_{zy}=0$,
the ($xy$) components of the tensor $\hat{\rho}$ obtained from (\ref{eq_res0}) by inverting $\hat{\mathcal{D}}$-matrix are decoupled from $\mathcal{D}_{zz}$.
The calculation of $\rho_{xx}$ for ${\bf B} \parallel {\bf e}_y$, on the contrary, generally requires the knowledge of other components $\mathcal{D}_{zx},\mathcal{D}_{zy}$.
Let us consider the $\theta$-independent EOD scattering model $\tau_s = {\rm const}(\theta)$ and
assume that the scattering time on short-range impurities is long compared with all other times $\tau_i \gg \tau_s, \omega_c^{-1}$.
In this case the diffusion coefficient in ${\bf B} \parallel {\bf e}_z$ geometry takes the form
$$\mathcal{D} = (1- \mathcal{P})  \mathcal{D}_0(\omega_c) + \mathcal{P} \mathcal{D}_{c},$$
where $\mathcal{P} = \exp{(- 2\pi/\omega_c \tau_s)}$.
The first term is due to
the contribution of $(1-\mathcal{P})$ "wandering" electrons described by the
ordinary diffusion coefficient $\mathcal{D}_0(\omega_c) = \langle v_x^2 \tau_s/(1 - i \omega_c \tau_s) \rangle$
at finite magnetic field. 
The second term is due to the pure Hall current
of the remaining $\mathcal{P}$ "circling" electrons
described by
$\mathcal{D}_c =  \langle v_x^2 \rangle/ (-i \omega_c)  $.
Inverting $\hat{\mathcal{D}}$-matrix gives the following expression for the resisitivity
\begin{equation}
    \label{eq_res_mem0}
    \rho_{xx}(B) \approx \rho_0 \left( 1 - \mathcal{P} \right),
    \qquad
    \rho_{0} = \frac{1}{e^2 \nu_F \mathcal{D}_0(\omega_c=0)}.
\end{equation}
This result naturally coincides with the resistivity expression obtained in~[18] for 2D systems.

\begin{figure}[t]
\centerline{\includegraphics[scale=0.8]{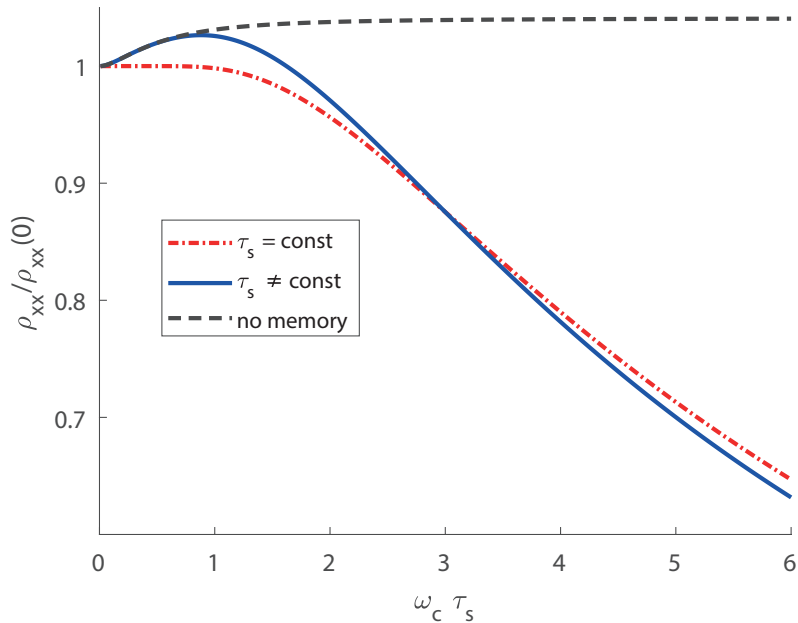}}
\caption{Comparison of isotropic (red dash-dotted line) and \\ $\theta$-dependent (blue solid line) models at $\tau_i/\tau_s = 10^{20}$. The black \\ dashed line is $\theta$-dependent model implementation without me- \\ mory effects (${\cal P}$ is set to be equal zero).}
\label{fig-TI-inf}
\end{figure}

Let us draw the attention to the most prominent feature of the considered memory effect mechanism.
We note that the magnetoresistance in ${\bf B} \parallel {\bf e}_y$ geometry is very weak (see Fig.~\ref{fig:rhoBzBy}), implying $\rho_{xx} \approx \rho_0$.
Therefore, the anisotropy of magnetoresistance acquires an exponential dependence on the magnetic field
\begin{equation}
    \label{DeltaAM_and_P}
    \Delta_{AM} \approx \mathcal{P}  = \exp{(- 2\pi/\omega_c \tau_s)}.
\end{equation}
We conclude that the magnitude of the effect increases with the increase of the magnetic field and becomes significantly important at $\omega_c \tau_s \ge 1$.
For instance, the magnetoresistane achieves $25\%$ at $\omega_c \tau_s \approx 3$, see Fig.~\ref{fig-TI-inf}.

Considering $\theta$-dependent models of the electron scattering on EOD does not noticeably change this situation.
Indeed, the expression for the diffusion coefficient in the limit $\tau_i/\tau_s \to \infty$ can be presented as
\begin{equation}
\mathcal{D} =
\Bigl\langle v_x^2 \tau_s
\frac{1 - \mathcal{P}_\theta}{1 - i \omega_c \tau_s}
\Bigr\rangle
+ \langle
v_x^2 \mathcal{P}_\theta
\rangle
\frac{1}{- i \omega_c}.
\end{equation}
The difference with the two-dimensional case and $\theta$-independent model lies in the fact, that the relative number of the "circling" electrons $\mathcal{P}_\theta = \exp{(-2\pi/\omega_c \tau_s(\theta))}$ now depends on the electron velocity projection $v_z = v_F \cos{{\theta}}$, hence the averaging over different $\theta$-electrons.
The calculated resisitivity $\rho_{xx}(B)$ is shown in Fig.~\ref{fig-TI-inf}.
The red and blue curves in Fig.~\ref{fig-TI-inf}
correspond to the isotropic $\tau_s = {\rm const}(\theta)$ and $\theta$-dependent $1/\tau_s = \sin{\theta}/\tau_s^0$ models, respectively,
the green curve demonstrates the resistivity of the $\theta$-dependent model without accounting for the memory effect.
The AMR-parameter for the $\theta$-dependent model can be estimated as the difference between the curves with and without accounting for the memory effect.

\begin{figure}[t]
\centerline{\includegraphics[scale=0.8]{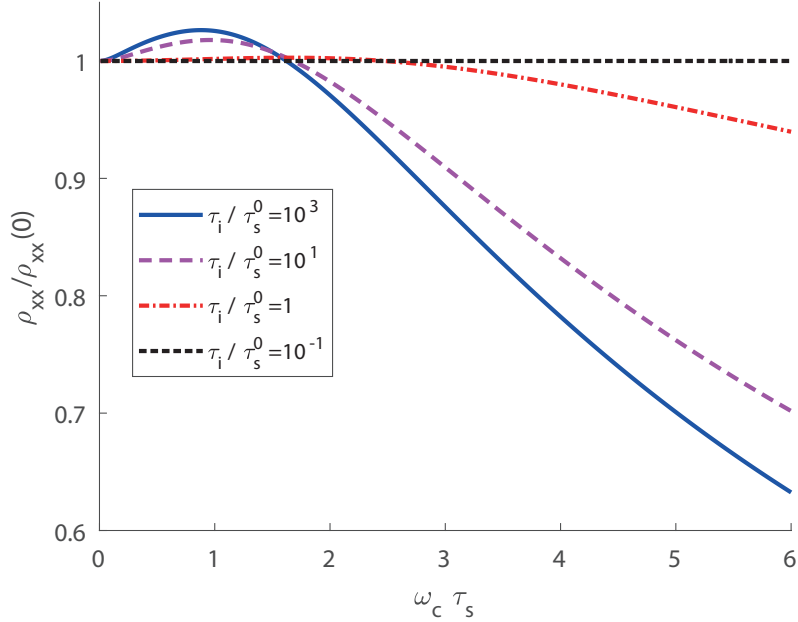}}
\caption{The memory effect at finite $\tau_i$ in nonisotropic ($\theta$-depen- \\ dent) model of scattering on EOD ($\tau_s^{-1} = (\tau_{s}^0)^{-1} \cdot \sin{{\theta}}$).}
\label{fig-NONisotr-TI}
\end{figure}

It is worth emphasizing that when both mechanisms of the magnetoresistance are presented,
the magnetoresistance in ${\bf B} \parallel {\bf e}_z$ geometry becomes sign-alternating, see the blue line in Fig.~\ref{fig-TI-inf}.
This happens because of the exponentially small number of "circling" electrons at small $\omega_c \tau_s \ll 1$.
In this case, the initial region of $\rho_{xx}(B)$ dependence remains dominated by the drift velocity spreading of the "wandering" electrons, while
the memory effect becomes active only at larger magnetic fields, when the number of "circling" electrons becomes noticeable enough.
The effect of the sign-alternative magnetoresistance is also an important property of bulk systems with the extended defects.

We further take into account the effect of the electron scattering on the short-range impurities.
In the limit $\tau_i \ll \tau_s $, opposite to the considered above,
the expression for the $\mathcal{D}$ turns into the ordinary diffusion coefficient
$\mathcal{D} = \langle v_x^2 \rangle \tau_i/(1-i\omega_c \tau_i)$
for an electron gas in the the external magnetic, here the single parameter $\tau_i$ describes the scattering on short-range impurities.
In this limit, naturally, there is no talking about the formation of "circling" electrons,
as the electron encounters EOD much rarely than the short-range impurities,
hence the memory effect becomes suppressed.
Fig.~\ref{fig-NONisotr-TI} demonstrates the behavior of $\rho_{xx}(B)$
for different values of the ratio $\tau_i/\tau_s^0$; here we employed the $\theta$-dependent model
$1/\tau_s = \sin{\theta}/\tau_s^0$ of the electron scattering on EOD.
As follows from Fig.~\ref{fig-NONisotr-TI}, the memory-effect induced magnetoresistance looses its key role only when
the scattering time on scalar impurities becomes comparable with that on EOD ($\tau_i \approx \tau_s$).
It is right up to this moment that the magnetoresistance is dominated in strong fields by the contribution due to "circling" electrons.
For instance,
for $\tau_i/\tau_s^0 = 10$
the magnitude of the effect at $\omega_c \tau_s \approx 4$ is still large,
approaching $20\%$.



%


\subsection{Discussion of effects from tilted magnetic fields and rosette trajectories}

It is of interest to study the effect of a weak tilting of the magnetic
field $\mathbf{B}$ from the EOD direction $\mathbf{e}_z$. Let us
denote the angle between these directions as $\alpha \ll 1$, the mean
distance between the EOD as $L=n_s^{-1/2}$ and the diameter of
EOD as $d \propto \sqrt{\sigma_s}$ (the diameter corresponds to the scattering cross-section $\sigma_s$). As for the considered above cases $\mathbf{B} ||
\mathbf{e}_z$ and $\mathbf{B} \perp \mathbf{e}_z$, we imply that the
typical magnitudes of the magnetic field $B$ correspond to the cyclotron
radius $R_c = v_\perp/\omega_c$ in the interval $ L \ll R_c \ll l_s$.
Here $ v_\perp = v_F \sin \theta $ is the absolute value of the
projection of the electron velocity on the $xy$ plane (defined as in the above sections) and $ l_s =
v_\perp \tau_s \sim L^2/d $ is the scattering length
corresponding to the velocity components in the $xy$ plane due to collisions with EOD.

The main effect of a small tilting of the magnetic field is the
appearance of very rare collisions of the electrons on the circling
orbits with EOD due to a slow drift of the orbit center in the
$xy$-plane. The wandering electrons are weakly affected by a deviation
of $\mathbf{B}$ from the EOD direction $z$. The tilt of the magnetic field can be considered small when
the shift of the center of circling electron trajectory projection
on the $xy$ plane after one
cyclotron period is small compared to
the distance between EOD, i.e.
$\alpha R_c \ll L $. Such drift of electron cyclotron circles in the
$xy$ plane 
results in collision of each circling electron with a
EOD after a certain number of rotations. This number can be
estimated as $N^*/p^*$, where $ N^*$ is the number of rotations by which
the electron center makes the distance $L$, $ N^* = L /(\alpha R_c )$,
and $ p^*$ is the probability to collide with a EOD when an electron
moves on the trajectory segment corresponding to one cyclotron period
and located on the closest distance from a EOD (as compared  with
other segments).
This probability can be estimated as $p^* \approx 1$ for very small tilting angles $\alpha R_c \ll d $, while for the angles satisfying $ d \ll \alpha R_c  \ll L$ it takes $p^* \approx d/(\alpha R_c) $.

In this way, we arrive to the estimate for the scattering time of the
in-plane velocity components $v_{x,y}$  of the ``almost circling
electrons'' due to rare collisions with the EOD $\tau^* \sim N^* T_c
/p^* $. In the case $\alpha R_c \ll d $ we obtain $\tau^*  \sim
L/(\alpha v_\perp ) $, whereas at $ d \ll \alpha R_c  \ll L$ we have $
\tau^*  \sim   L /(d \omega_c)$. The described effect of the rare
collisions of the circling electrons with EOD plays the role in the
electron dynamics analogous to the scattering of the circling electrons
on short-range impurities in the bulk, thus it also determines the
conditional probability $\tilde{P}'  (\tau_1|T_c)$ by the similar way as
scattering on the bulk impurities.
Thus, the equation for the diffusion coefficients in (\ref{eq_D-memory})
should be modified {by changing $1/\tau_i$ by $1/\tau_i + 1/\tau^* (\theta)$} in order to {take into} account both scattering mechanisms. 

It is worth noting that in the leading order with respect to $\mathcal{P}$ there is another microscopic mechanism for the memory effect.
Namely, upon the external magnetic field the electron moving along the cyclotron circle
has a finite probability to encounter the same scatterer many times in a row.
This memory effect results in the formation of the so-called rosette trajectories, which effectively lead to the localization of the electron in real space~[21,22].
Our analysis based on~[21,22] shows that, despite the magnitude of this phenomenon is controlled by the same parameter $\mathcal{P} = e^{-2\pi/\omega_c \tau}$ as the appearance of the "circling" electrons,
the main effect of the rosette-induced localization concerns the renormalization of the transport time~[21] and it does not critically affect the effect of negative magnetoresistance, considered in detail in this paper.

Both these two groups of effects, the magnetoresistanstance at tilted magnetic fields and the extended collisions (rosettes), require further studies in order to obtained the details of the picture presented in this section.


\section{Conclusion}

We demonstrated that in bulk non-magnetic materials with an isotropic Fermi surface and extended one-dimensional defects,
such as skrew dislocations or charge stripes, the strong magnetoresistance effect (up to $100\%$ change) arises when the magnetic field
becomes co-aligned with the defect axis.
This feature stems from the memory effect, which is the existence
of collisionless electron trajectories in a static pattern of EOD.
Indeed,
upon the external magnetic field
there is always a finite number of electrons
which do not experience the scattering on EOD during one cycloton period.
Essentially, when the magnetic field is directed along EOD-axis these electrons will keep moving along the spiral trajectories
winding around EOD without any subsequent
scattering on them.
This memory effect results in the enhanced negative magnetoresistance in the plane perpendicular to EOD-axis.
Applying the magnetic field in the transverse direction (perpendicular to EOD-axis) destroys the trajectory correlations responsible for the memory effect,
thus reducing the magnitude of the magnetoresistance.
We analyzed the electric kinetics in this geometry and demonstrated that
the magnetoresistivity still can arise from the velocity direction dependence of the microscopic electron scattering rates on EOD.
However, the obtained magnitude of the effect does not exceed several percents, thus indicating the appearance of the strong anisotropy of the magnetoresistance.
Based on 
the correlation function approach we derived
the general analytical expression
which describes the electrical resistivity with account for the memory effect.
The noteworthy advantage of the developed theory is that it allowed us to take into account the electron scattering on short-range impurities and the deviation of the magnetic field direction from the EOD-axis as well.
Both this factors result in the suppression of the magnetoresistance anisotropy.
We demonstrated that the magnetoresistance in classicaly strong magnetic fields acquires an exponential character
and it reaches large values for $\omega_c \tau_s \approx 1$, where $\omega_c$ is a cyclotron frequency and $\tau_s$ is a scattering time on 1D defects.
We found out that the memory effect gets suppressed when the scattering times corresponding to EOD and short-range impurities become comparable.

\ack
We thank A.P.~Dmitriev from Ioffe Institute and D.G.~Polyakov from Karlsruhe Institute of Technology for fruitful discussions. This work was supported by Russian Science Foundation (analytical theory -- project No. 18-72-10111) and by Russian Foundation for Basic Research (numerical calculations -- project No. 18-02-01016).
\\
\\
\\
{\bf { \Large References}} 
\\
\\
\noindent
[1] McGuire T and Potter R 1975 IEEE Transactions on Magnetics 11 1018-1038

\noindent
[2] Wadehra N, Tomar R, Varma R M, Gopal R K, Singh Y, Dattagupta S and Chakraverty S 2020
Nature Communications 11 847

\noindent
[3] Liu M, Hong Y, Xue H, Meng J, Jiang W, Zhang Z, Ling J, Dou R, Xiong C, He L and Nie J 2020
Journal of Physics: Condensed Matter 32 235003

\noindent
[4] Burgos R, Warnes J and De La Espriella N 2018 Journal of Magnetism and Magnetic Materials
466 234-237

\noindent
[5] Du W, Tang X, Liu M, Su H, Han G and Lu Q 2020 Journal of Physics: Condensed Matter 32
235802

\noindent
[6] Miller B and Dahlberg E D 1996 Applied physics letters 69 3932-3934

\noindent
[7] Gredig T, Krivorotov I N and Dahlberg E D 2002 Journal of applied physics 91 7760-7762

\noindent
[8] Pavlosiuk O, Kleinert M, Swatek P, Kaczorowski D and Winiewski P 2017 Scientific Reports 7
12822

\noindent
[9] Novak M, Zhang S N, Orbanic F, Biliskov N, Eguchi G, Paschen S, Kimura A, Wang X X, Osada
T, Uchida K, Sato M, Wu Q S, Yazyev O V and Kokanovic I 2019 Phys. Rev. B 100(8) 085137

\noindent
[10] Krasikov K M, Azarevich A N, Glushkov V V, Demishev S V, Khoroshilov A L, Bog

\noindent
[11] Zhang K, Du Y, Wang P, Wei L, Li L, Zhang Q, Qin W, Lin Z, Cheng B, Wang Y, Xu H, Fan X,
Sun Z, Wan X and Zeng C 2020 Chinese Physics Letters 37 090301

\noindent
[12] Yan J, Luo X, Gao J J, Lv H Y, Xi C Y, Sun Y, Lu W J, Tong P, Sheng Z G, Zhu X B, Song
W H and Sun Y P 2020 Journal of Physics: Condensed Matter 32 315702

\noindent
[13] Morimoto T and Nagaosa N 2016 Phys. Rev. Lett. 117(14) 146603

\noindent
[14] Jho Y S and Kim K S 2013 Phys. Rev. B 87(20) 205133

\noindent
[15] Aggarwal N, Krishna S, Goswami L and Gupta G 2021 Materials Science and Engineering: B 263
114879

\noindent
[16] Kim G, Feng B, Ryu S, Cho H J, Jeen H, Ikuhara Y and Ohta H 2021 ACS Appl. Mater. Interfaces
13(5) 68646869

\noindent
[17] Baskin E M, Magarill L N and Entin M 1978 Sov. Phys. JETP 48 365

\noindent
[18] Dmitriev A, Dyakonov M and Jullien R 2001 Phys. Rev. B 64(23) 233321

\noindent
[19] Dmitriev I A, Mirlin A D, Polyakov D G and Zudov M A 2012 Rev. Mod. Phys. 84(4) 1709-1763

\noindent
[20] Bobylev A V, Maao F A, Hansen A and Hauge E H 1995 Phys. Rev. Lett. 75(2) 197-200

\noindent
[21] Bobylev A V, Maao F A, Hansen A and Hauge E H 1997 J. Stat. Phys. 87 12051228

\noindent
[22] Beltukov Y M and Dyakonov M I 2016 Phys. Rev. Lett. 116(17) 176801

\noindent
[23] Chepelianskii A D and Shepelyansky D L 2018 Phys. Rev. B 97(12) 125415

\noindent
[24] Alekseev P S and Alekseeva A P private communication

\noindent
[25] Murzin S S 1984 JETP Lett. 39 695

\noindent
[26] Polyakov D 1986 Sov. Phys. JETP 63 317

\noindent
[27] Bonch-Bruevich V L and Kalashnikov S G 1977 Semiconductors physics (Moscow Editions Nauka)

\noindent
[28] Averkiev N and Dyakonov M 1982 JETP Letters 35 196-198

\noindent
[29] Dong S H and Lemus R 2002 Applied Mathematics Letters 15 541-546



\appendix

\section{Microscopic derivation of the collision integral with 1D defects}
\label{App-CollisionIntegral}
The collision integral entering
in the Boltzmann kinetic equation
reads for elastic electron scattering processes as
\begin{equation}
\label{collision-integral}
    {\rm St}[f_{\bf p}] = \sum\limits_{\bf p'} \left( W_{\bf p, p'} f_{\bf p'} - W_{\bf p', p} f_{\bf p} \right) ,
\end{equation}
where $f_{\bf p}$ is the distribution function of electrons with momentum ${\bf p}$ and
$W_{\bf p,p'}$ is the scattering rate corresponding to the transition from $\bf p'$ to $\bf p$ states.
According to the Fermi golden rule $W_{\bf p,p'}$ is given by
\begin{equation}
\label{Wpp}
    W_{\bf p, p'} = \frac{N_{3D}}{\Omega} \frac{2\pi}{\hbar} |V_{\bf p, p'}|^2 \delta (\varepsilon_{\bf p} - \varepsilon_{\bf p'}) .
\end{equation}
Here $\Omega$ is the volume of the sample and $N_{3D}$ is a number of the scatterers in this volume. The scattering matrix element $V_{\bf p, p'}$
in case of 1D defects in 3D material takes into account the momentum conservation along the $\bm{e}_z$ axis of these defects
\begin{equation}
\label{Vpp}
    V_{\bf p, p'}
    = u_{\parallel}({\bf q}) 2\pi \delta(k_z - k_z').
\end{equation}
Here we introduced the transversal scattering matrix element $u_{\parallel}({\bf q})$ depending on the perpendicular to EOD components of wave vector change ${\bf q} \equiv \bm{k}_{\perp} - \bm{k}_{\perp}'$. 
Taking the square of the matrix element we use a standard change $\delta^2(k_z) \rightarrow \delta(k_z) L_z/2\pi$ introducing the linear size of the system along EOD direction $L_z$. Then the scattering integral takes the following form
\begin{equation}
\label{STfp_with_delta}
     {\rm St}\left[f_{{\bf p}}\right] = \frac{2\pi}{\hbar} \nu_3 n_2 \int \frac{d\Omega'}{4\pi} 2\pi \delta(k_z - k_z')
    |u(\theta; \varphi-\varphi')|^2
     \left(
    f_{\bf{p}'}-f_{\bf{p}}
     \right),
\end{equation}
where $(\theta,\varphi)$ are the angular coordinates of the momentum $\bm{p}$,
$\nu_3 = g_3(E_F)/\Omega$ is the density of states on a Fermi level and $n_2 = n_3 L_z$ is a 2D concentration of 1D defects.
Let us further introduce an effective 2D electron differential scattering cross-section as $d\sigma/d\varphi = \nu_2 |u|^2/\hbar v_{\perp}$, where $\nu_2 = m/2\pi \hbar^2$ is the 2D density of states.
Using $d\sigma /d \varphi$ we express the collision integral as follows
\begin{equation}
\label{STfp_with_sigma}
    {\rm St}\left[f_{\bm p}\right] =
    \frac{\tilde{\nu_2}}{\nu_2} n_2 v_\perp \int d\varphi' \left( \frac{d\sigma}{d\varphi'} f_{\bf{p}'} - \frac{d\sigma}{d\varphi'} f_{\bf{p}} \right),
\end{equation}
where 
$\tilde{\nu_2} = {\pi}\nu_3/{k_F} $ is an effective 2D density of states.
For the $(xy)$-plane isotropic scattering model implied in our paper
the cross-section $d\sigma/d\varphi$ does not depend on the polar scattering angle,
which brings us to the collision integral in the following form:
\begin{equation}
    {\rm St}[f_p] =  - \frac{\delta f_p - \langle \delta f_p \rangle_\varphi}{\tau_s},
    \qquad
    \frac{1}{\tau_s} = \frac{\tilde{\nu}_2}{\nu_2} n_2 v_{\perp} \sigma_s,
\end{equation}
where $\sigma_s$ is the total scattering cross-section.


\section{Numerical solution of the Boltzmann equation with the account for an angle-dependent scattering on 1D defects}
\label{App-NumSol-Boltzmann}

We are interested in solving the kinetic Boltzmann equation in ${\bf B} \parallel {\bf e}_y$ geometry, when
scattering time $\tau_s$ depends on incident velocities angles of electrons and
the analytical solution is not possible. 
We consider
the model when $1/\tau_s = (1/\tau_s^0) \sin{\theta}$, where $\tau_s^0$ is some constant. Starting from (\ref{eq-kinetic}) and looking for its solution as a decomposition in spherical harmonics $\delta f_p = \sum_{l,m} f_{l,m} Y^m_l (\theta,\varphi)$ one can obtain
\begin{eqnarray}
\label{Spherical_harmonics_kinetic_eq}
\fl
    \alpha q_{\gamma} + \frac{\omega_c}{2} \sum\limits_{l,m} f_{l,m} \left[ \sqrt{(l-m)(l+m+1)} Y^{m+1}_l - \sqrt{(l+m)(l-m+1)} Y^{m-1}_l \right] = \nonumber \\
    = - \sum\limits_{l,m} f_{l,m} \left[ \frac{1}{\tau_i} Y^m_l + (1-\delta_{m,0}) \frac{1}{\tau^0_s} \sin{\theta}~\! Y^m_l  \right],
\end{eqnarray}
where $\alpha=(eE_0/m\sqrt{2})\partial f_p^0 / \partial v$ and ${\bf E} = E_0 {\bf e}_{\gamma}$ with $\gamma = x,y,z$ resulting in appropriate combinations of angular harmonics $q_{\gamma}$ (for example, $q_x = Y_1^{-1} - Y_1^1$). 
Multiplying both left and right hand sides of (\ref{Spherical_harmonics_kinetic_eq}) by complex conjugate of some $Y^{q}_k$ and integrating over the solid angle
we get
\begin{eqnarray}
\label{Spherical_harmonics_algebraic_eq}
\fl
    f_{k,q-1} \frac{\omega_c}{2} \sqrt{(k-q+1)(k+q)} - f_{k,q+1} \frac{\omega_c}{2} \sqrt{(k+q+1)(k-q)} +
    f_{k,q} \frac{1}{\tau_i} + \nonumber \\
    + \sum\limits_{l} f_{l,q} (1-\delta_{q,0}) \frac{\beta_{l,k}^{q}}{\tau^0_s} \int\limits_{-1}^{1} dx P_k^{q} (x) P_1^1 (x) P_l^{q} (x) = - \alpha \delta_{k,1} (\delta_{q,-1} - \delta_{q,1}),
\end{eqnarray}
which is an algebraic system of linear equations on $f_{k,q}$. Here we have used a concrete choice of ${\bf E} = E_0 {\bf e}_{x}$, resulting in a specific form of the right hand side.
For other directions of electric field the right hand side will change, though it is always determined by
$k=1$ harmonics linear combinations. The coefficient $\beta$ stems from the spherical harmonics normalization conditions and the fact that $\sin{\theta} = - P_1^1 (\cos{\theta})$, thus
\begin{eqnarray}
\label{beta_coeff}
    \beta_{l,k}^{q} = -\frac{1}{2} \sqrt{(2l+1)(2k+1)} \sqrt{\frac{(l-q)!}{(l+q)!} \frac{(k-q)!}{(k+q)!}}.
\end{eqnarray}
The latter is defined for positive $q$'s and it should give us a unit when multiplied by the integral of associated Legendre polynomials as in (\ref{Spherical_harmonics_algebraic_eq}), but where the middle one polynomial $-P_1^1 (x)$ is eliminated and $k$ equals $l$.

One can see that the first three terms in (\ref{Spherical_harmonics_algebraic_eq}) couple coefficients $f_{k,q}$ with different $q$ (with their $\pm 1$ change), but leave $k$ the same. On the contrast, the last term in left hand side leaves the $q$ index fixed, while couples the coefficients $f_{k,q}$ with different $k$. The integral in the last term, which is an overlap integral of three associated Legendre polynomials, could be computed using the formula and the algorithm described in~[29], and one can prove that there is an infinite number of harmonics that couple to the $k=1$ one. These harmonics are $l=1, 3, 5, 7, \dots$ and etc.

When $q=0$ the last term in left hand side of (\ref{Spherical_harmonics_algebraic_eq}) is zero and the system decomposes to rather simple relations
\begin{eqnarray}
\label{fk0}
    f_{k,0} = \frac{\omega_c \tau_i }{2} \sqrt{k(k+1)} (f_{k,1} - f_{k,-1}) .
\end{eqnarray}
This is due to the fact that the electrons moving along the 1D defects axis are not scattered by them,
so the relaxation  of their velocities is entirely due to short-range impurities described by $\tau_i$.

Nevertheless, in the case $q \neq 0$, different $f_{l,q}$ start to affect the value of $f_{1,q}$. This means that the real system (\ref{Spherical_harmonics_algebraic_eq}) is infinite. To conduct numerical calculations we set $l_{max} = 15$, although the changes in calculation results become negligible starting even from $l=7$ harmonics accounting.

We have rearranged a second order tensor $f_{k,q}$ to a vector form
in a following manner $[f_{1,-1},f_{1,0},f_{1,1},f_{3,-3},f_{3,-2},\dots]^T$. This obvious rule for $f_{k,q}$ listing implies a consequent enumeration of all $k=1,3,5,7,\dots,l_{max}$ (as soon as only these harmonics influence the $k=1$ one, in which we are interested) with an appropriate account for corresponding $q$ values changing from $q=-k$ to $q=k$. The general formula for a new composite vector index is $n=k+q+1+k(k-1)/2$, which takes the values $n=1,\dots,N$, where $N=(l_{max}+1)(l_{max}+2)/2$. Thus, (\ref{Spherical_harmonics_algebraic_eq}) is transformed to a linear algebraic equation in a matrix form of $N\times N$ dimension with inhomogeneous part in the right hand side proportional to $\alpha$. $N=136$ in our case and it grows with a square rate at the $l_{max}$ increase, but note that an appropriate matrix of linear coefficients contains a lot of zeros.

Solving this linear system of equations we get numerical values of $f_{1,-1},f_{1,0},f_{1,1}$, which determine all the components of the diffusion tensor. For example, $\mathcal{D}_{xx} \propto  (f_{1,-1} - f_{1,1})/\sqrt{2}$, $\mathcal{D}_{yx} \propto  (-i)(f_{1,-1} + f_{1,1})/\sqrt{2}$ and $\mathcal{D}_{zx} \propto  f_{1,0}$, when the ${\bf E} \parallel {\bf e}_{x}$ condition determines (\ref{Spherical_harmonics_algebraic_eq}) in its right hand side. To calculate other components we need to repeat calculations with right hand side of (\ref{Spherical_harmonics_algebraic_eq}) related to ${\bf E} \parallel {\bf e}_{y}$ and ${\bf E} \parallel {\bf e}_{z}$. 


\section{Calculation of the diffusion coefficient $\mathcal{D}_{zz}$ within the correlation function approach}
\label{App-DiffCoef}

The calculation of $\mathcal{D}_{zz}$ by means of the correlation functions requires to take into account an infinite series of scattering events on EOD. Indeed, since the electron scattering on EOD does not lead to the relaxation of $v_z$,
it requires an electron to meet a short-range impurity to destroy the correlation with $v_z(0)$.
This, however, can happen after an arbitrary number of scattering acts on EOD.

\begin{figure}[t]
\centerline{\includegraphics[width=0.7\textwidth]{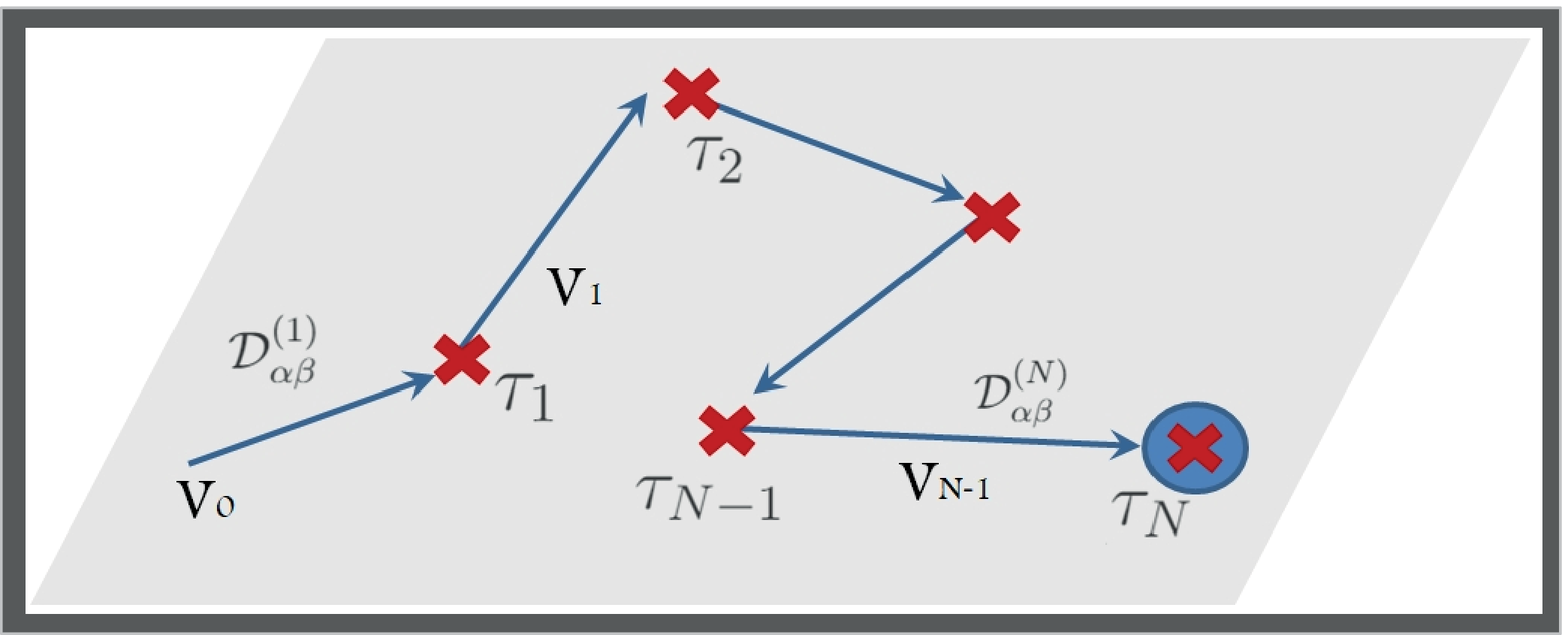}}
\caption{The electron diffusion trajectory.}
\label{fig:polygon}
\end{figure}

The trajectory of an electron with the initial velocity ${\bf v}(0)$ is the polygonal line with the kinks corresponding to the scattering events (see Fig.~\ref{fig:polygon}).
The segment ($N-1,N$) between neighboring kinks is the cyclotron arc whose orientation is determined by the velocity direction ${\bf v}_{N-1}$ after $(N-1)$-th collision.
The probability density regulating the location of kinks on each trajectory is determined by the Poisson stochastic process, see  (\ref{eq:Poisson}).
At the same moment,
the probability density to experience $N$ subsequent collisions each at time $t_i$ is simply given by the product of single scattering event probabilities; for instance $P_s(t_N-t_{N-1}) dt_N P_s(t_{N-1}-t_{N-2}) dt_{N-1} \dots P_s(t_1) dt_1$ describes the electron path for which the first $N$ collisions were due to the scattering on EOD.
We calculate the diffusion coefficient by summing separately the contributions to the time integral (\ref{eq_D-CFT})
stemming from different segments of the electron trajectories:
\begin{equation}
    \label{eq_Dif-Simple}
 \mathcal{D}_{\alpha \beta} =\sum_{N=1}^{\infty} \mathcal{D}_{\alpha \beta}^{(N)},
\end{equation}
where the $N$-th term $\mathcal{D}_{zz}^{(N)}$ collects
the correlation of the electron velocity with its value ${\bf v}(0)$ at the initial moment from all trajectories at the time interval between $N-1$ and $N$ scattering events.
Basically, such correlation in $\mathcal{D}_{zz}^{(N)}$ is preserved only
for such trajectories which have $(N-1)$ subsequent scattering acts on EODs.
The probability density to experience $(N-1)$ scattering events on EOD in a row with an arbitrary type of scattering at $t_N$ time moment is given by $P_0(t_N-t_{N-1}) dt_N P_s(t_{N-1} - t_{N-2})dt_{N-1} \dots P_s(t_1) dt_1$, the corresponding contribution to the diffusion coefficient reads as
\begin{equation}
\fl
    \mathcal{D}_{zz}^{(N)} =
    \Bigl\langle
    \int\limits_0^{\infty} d\tau_1 P_s(\tau_1) \dots
    \int\limits_{\tau_{N-2}}^{\infty} d\tau_{N-1} P_s(\tau_{N-1})
    \int\limits_{\tau_{N-1}}^{\infty} d\tau_N P_0(\tau_N)
    \int\limits_{\tau_{N-1}}^{\tau_N} dt \cdot v_z(t) v_z(0)
    \Bigr\rangle,
\end{equation}
here $\tau_k=t_k - t_{k-1}$, $k=2 \dots N$ ($\tau_1=t_1$) and the average $\langle \dots \rangle$ goes over the electron isotropic initial velocity direction distribution.
Since $v_z(t) = v_z(0)$ (until the electron scatters a short-range defect), the series for the diffusion coefficient transforms into the geometric progression, which can be easily evaluated
\begin{equation}
\mathcal{D}_{zz}
= \Bigl\langle
v_z^2 \tau_0
\sum_{n=0} \left( \frac{\tau_0}{\tau_s} \right)^n
\Bigr\rangle
= \langle v_z^2 \tau_i \rangle.
\end{equation}
Naturally, we obtained essentially the same expressions for the diffusion coefficients ${\mathcal{D}_{zz}}$ as we did using the kinetic equation approach, see (\ref{eq_D-A-Kin}).
%
%
%

\end{document}